\documentclass[10pt,journal,compsoc]{IEEEtran}
\usepackage{graphicx}
\usepackage{ifpdf}
\usepackage{balance}
\usepackage{times}
\usepackage{comment}
\usepackage{xcolor}
\usepackage{soul}
\usepackage[utf8]{inputenc}
\usepackage[small]{caption}
\usepackage{amsfonts}
\usepackage{amsmath}
\usepackage{epsfig}
\usepackage{amsopn}
\usepackage{float}
\usepackage{epstopdf}

\usepackage{url}
\usepackage{booktabs} % For formal tables
\usepackage{algorithmic}
\usepackage{colortbl}
\usepackage[ruled]{algorithm2e} % For algorithms

\usepackage{caption}

\usepackage{multirow}
\usepackage{tabularx}
\usepackage{bm}
\usepackage{graphicx}
\usepackage{rotating}
\usepackage{mathrsfs}

\ifCLASSOPTIONcompsoc
  \usepackage[nocompress]{cite}
\else
  \usepackage{cite}
\fi

\usepackage{amsmath}
\usepackage{algorithmic}

\usepackage{array}

\ifCLASSOPTIONcompsoc
 \usepackage[caption=false,font=footnotesize,labelfont=sf,textfont=sf]{subfig}
\else
 \usepackage[caption=false,font=footnotesize]{subfig}
\fi

\usepackage{url}
\begin{document}
\title{Internet of Things Meets Brain-Computer Interface: A Unified Deep Learning Framework for Enabling Human-Thing Cognitive Interactivity}

\author{Xiang~Zhang,~\IEEEmembership{Student Member,~IEEE,}
        Lina~Yao,~\IEEEmembership{Member,~IEEE,}
        Shuai~Zhang,~\IEEEmembership{Student Member,~IEEE,}\\
        Salil Kanhere,~\IEEEmembership{Member,~IEEE,}
        Michael Sheng,~\IEEEmembership{Member,~IEEE,}
        and~Yunhao~Liu,~\IEEEmembership{Fellow,~IEEE}% <-this % stops a space
\IEEEcompsocitemizethanks{
\IEEEcompsocthanksitem Xiang Zhang, 
Lina Yao, Shuai Zhang, Salil Kanhere  are with the School of Computer Science and Engineering, University of New South Wales, Kensington, NSW 2052, Australia.\protect\\
E-mail: {xiang.zhang3, shuai.zhang}@student.unsw.edu.au\protect\\
\{lina.yao, salil.kanhere\}@unsw.edu.au
\IEEEcompsocthanksitem Michael Sheng is with the Department of Computing, Macquarie University, Sydney, NSW 2109, Australia. E-mail: michael.sheng@mq.edu.au
\IEEEcompsocthanksitem Yunhao Liu is with the Department of Computer Science and Engineering, Michigan State University, MI 48824, USA. E-mail: yunhao@cse.msu.edu

}
\thanks{Copyright (c) 2012 IEEE. Personal use of this material is permitted. However, permission to use this material for any other purposes must be obtained from the IEEE by sending a request to pubs-permissions@ieee.org.}
}

% The paper headers
\markboth{Journal of \LaTeX\ Class Files,~Vol.~14, No.~8, August~2015}%
{Xiang \MakeLowercase{\textit{et al.}}: Internet of Things Meets Brain-Computer Interface}

% use for special paper notices
%\IEEEspecialpapernotice{(Invited Paper)}

\IEEEtitleabstractindextext{%
\begin{abstract}
A Brain-Computer Interface (BCI) acquires brain signals, 
%analyses them, 
analyzes and translates them into commands that are relayed to actuation devices %that 
for carrying out desired actions. 
With the widespread connectivity of everyday devices realized by the advent of the Internet of Things (IoT), BCI can empower individuals to directly control objects such as smart home appliances or assistive robots, directly via their thoughts. However, realization of this vision is faced with a number of challenges, most importantly being the issue of accurately interpreting the intent of the individual from the raw brain signals 
%which 
that are often of low fidelity and subject to noise. Moreover, pre-processing 
%the 
brain signals and the subsequent feature engineering are both time-consuming and highly reliant on human domain expertise. To address the aforementioned issues, in this paper, we propose a unified deep learning based framework that enables effective human-thing cognitive interactivity in order to bridge individuals and IoT objects. We design a reinforcement learning based Selective Attention Mechanism (SAM) to discover the distinctive features from the input brain signals. In addition, we propose a modified Long Short-Term Memory (LSTM) to distinguish the inter-dimensional information forwarded from the SAM. To evaluate the efficiency of the proposed framework, we conduct extensive real-world experiments and demonstrate that our model outperforms a number of competitive state-of-the-art baselines. Two practical real-time human-thing cognitive interaction applications are presented to 
%assess 
validate the feasibility of our approach.
% Electroencephalogram (EEG) 
\end{abstract}

}

% make the title area
\maketitle

% To allow for easy dual compilation without having to reenter the
% abstract/keywords data, the \IEEEtitleabstractindextext text will
% not be used in maketitle, but will appear (i.e., to be "transported")
% here as \IEEEdisplaynontitleabstractindextext when the compsoc 
% or transmag modes are not selected <OR> if conference mode is selected 
% - because all conference papers position the abstract like regular
% papers do.
\IEEEdisplaynontitleabstractindextext
% \IEEEdisplaynontitleabstractindextext has no effect when using
% compsoc or transmag under a non-conference mode.

% For peer review papers, you can put extra information on the cover
% page as needed:
% \ifCLASSOPTIONpeerreview
% \begin{center} \bfseries EDICS Category: 3-BBND \end{center}
% \fi
%
% For peerreview papers, this IEEEtran command inserts a page break and
% creates the second title. It will be ignored for other modes.
\IEEEpeerreviewmaketitle

\IEEEraisesectionheading{\section{Introduction}\label{sec:introduction}}

\IEEEPARstart{I}t is expected that by 2020 over 50 billion devices will be connected to the Internet. The proliferation of 
the 
Internet of Things (IoT) is expected to improve efficiency and impact various domains including home automation, manufacturing and industries, transportation and healthcare \cite{yao2015web}. 
%, yao-ic2015, Tran-csur17
%\textcolor{red}{add more citations}. 
Individuals will have the opportunity to interact and control a wide range of everyday objects through various means of interactions including applications running on their smartphone or wearable devices, voice and gestures. Brain-Computer Interface (BCI) \footnote{The BCI mentioned in this paper refers to non-invasive BCI.} is emerging as a novel alternative for supporting interaction bewteen IoT objects and individuals. BCI establishes a direct communication pathway between human brain and an external device thus eliminating the need for typical information delivery methods \cite{vallabhaneni2005brain}. Recent trends in BCI research have witnessed the translation of human thinking capabilities into physical actions such as mind-controlled wheelchairs and IoT-enabled appliances \cite{telesusing, jagadish2017novel}. These examples suggest that the BCI is going to be a major aiding technology in human-thing interaction \cite{zhang2017converting}. 

%BCI supports a secure, convenient, and instantaneous human-thing interactivity in IoT systems by empowering individuals to communicate with the outer world directly through their brain activities \cite{zhang2017converting}. 

BCI-based cognitive interactivity 
%affords 
offers several advantages. 
%First, 
One is the inherent privacy arising from the fact that brain activity is invisible and thus impossible to observe and replicate \cite{nguyen2017investigating}. %Second, 
The other is the convenience and real-time nature of the interaction, since the human only needs to think of the interaction rather than undertake 
%any 
the corresponding 
physical motions (e.g., speak, type, gesture) \cite{nakamura2018ear}.

%During the cognitive human-thing interactivity, the individual can communicate with the IoT devices only by \textit{thinking}, without any physical movements (e.g., speak, type). 

However, the BCI-based human-thing cognitive interactivity faces several challenges. While the brain signals can be measured using 
a number of technologies such as 
%EEG can be measured by 
Electroencephalogram (EEG) \cite{vallabhaneni2005brain}, Functional Near-Infrared Spectroscopy (fNIR) \cite{rahman2017evaluating}, and Magnetoencephalography (MEG) \cite{iijima2017cortical}, all of these methods are susceptible to low fidelity and are also easily influenced
by environmental factors and sentiment status (e.g., noise, concentration) \cite{rahman2016straight}. 
In other words, the brain signals generally have very low signal-to-noise ratios, and inherently lack sufficient spatial or temporal
resolution and insight on activities of deep brain structures \cite{zhang2017converting}. 
As a result, while current cognitive recognition systems can achieve about 70-80\% accuracy, this is not sufficient to design practical systems. Second, data pre-processing, parameter selection (e.g., filter type, filtering band, segment window, and overlapping), and feature engineering (e.g., feature selection and extraction both in the time domain and frequency domain) are all time-consuming and highly dependent on human expertise in the domain \cite{haselsteiner2000using}. 

\begin{figure}[!t]
\centering
\includegraphics[width=1\linewidth]{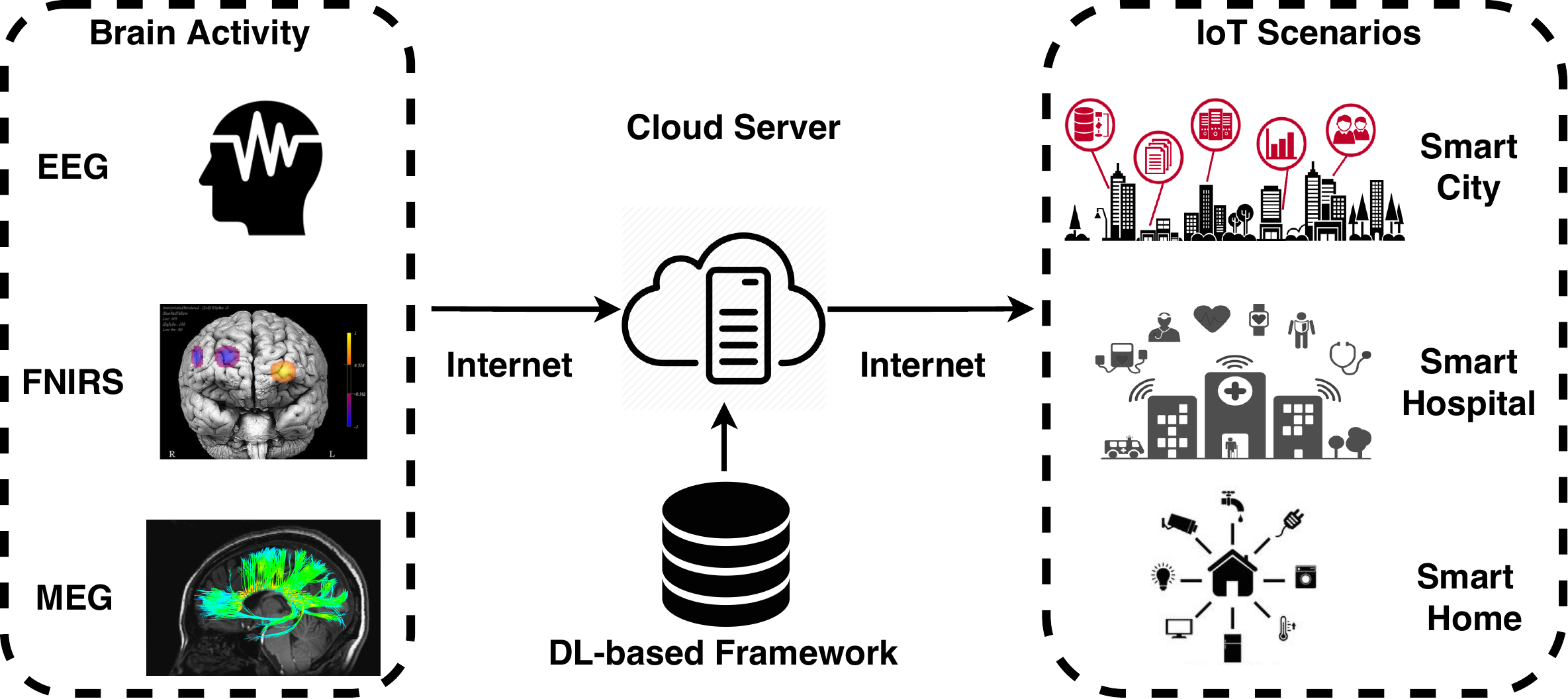}
\caption{Schematic diagram of cognitive IoT framework}
\label{fig_schematic}
\vspace{-5mm}
\end{figure}

To address the aforementioned issues, in this paper, we propose a unified Deep Learning (DL) \cite{lecun2015deep} framework for enabling human-thing cognitive interactivity. As shown in Figure~\ref{fig_schematic}, our framework measures the user's brain activity (such as EEG, FNIRS, and MEG) through 
%the 
a 
specific brain signal collection equipment. 
The raw brain signals are forwarded to the cloud server via Internet access. The cloud server uses a person-dependent pre-trained deep learning model for analyzing the raw signals.
% and analyzes the received brain signal. 
The analysis results interpreted signals could be used for actuating functions in a wide range of IoT applicants such as  smart city \cite{angelidou2017smart} (e.g., transportation control, agenda schedule), smart hospital \cite{dhariwal2017architecture,zhang2018multi} (e.g., emergency call, anomaly mentoring), and smart home \cite{al2017smart, yao2018wits} (e.g., appliances control, assistive robot control).

The proposed unified deep learning framework aims to interpret the subjects' intent and decode it into 
the 
corresponding commands which are discernible for the IoT devices. Based on our previous study \cite{zhang2017converting,zhang2017intent}, for each single brain signal sample, the self-similarity is always higher than the cross-similarity, which means that the intra-intent cohesion of the samples is stronger than the inter-intent cohesion. In this paper, we propose a weighted average spatial Long Short-Term Memory (WAS-LSTM) to exploit the latent correlation between signal dimensions. 
The proposed end-to-end framework is capable of modeling high-level, robust and salient feature representations hidden in the raw human brain signal streams and capturing complex relationships within data. 
The main contributions of this paper are highlighted as follows:
\begin{itemize}
  \item We propose a unified deep learning based framework to interpret individuals' brain activity for enabling human-thing cognitive interactivity. To our best knowledge, 
%michael: not a meaningful sentence!  
 we are the very first work that bridging 
% our work is the very first that bridges 
BCI and IoT to investigate end-to-end cognitive brain-to-thing interaction.
  \item We apply deep reinforcement learning, with designed reward, state, and action model, to automatically discover the most distinguishable features from the input brain signals. The discovered features are forwarded to a modified deep learning structure, in particular, the proposed WAS-LSTM, to capture the cross-dimensional dependency in order to recognize user's intention. 
  \item We also present two operational prototypes of the proposed framework: a brain typing system and a cognitive controlled smart home service robot, which demonstrate the efficacy and practicality of our approach.
\end{itemize}

% The rest of this paper is organized as follows. 
% Section~\ref{sec:cognition_detection_framework} describes the proposed cognition detection framework with all the functional units. Section~\ref{sec:experiments} extensively evaluates the performance of the proposed model.
% and reports the comparison with the state-of-the-art baselines. 
%  Section~\ref{sec:case_study} provides two case studies including a cognitive brain typing system and a cognitive robot in the smart home environment based on the proposed framework. Section~\ref{sec:discussion} outlines some discussion points and suggests future research directions, followed by Section~\ref{sec:conclusion} which concludes the paper. 

% \section{Related Work} % (fold)
% \label{sec:related_work}
% write this based on the residual space.

\begin{figure*}[!t]
\centering
  \includegraphics[width=0.8\linewidth]{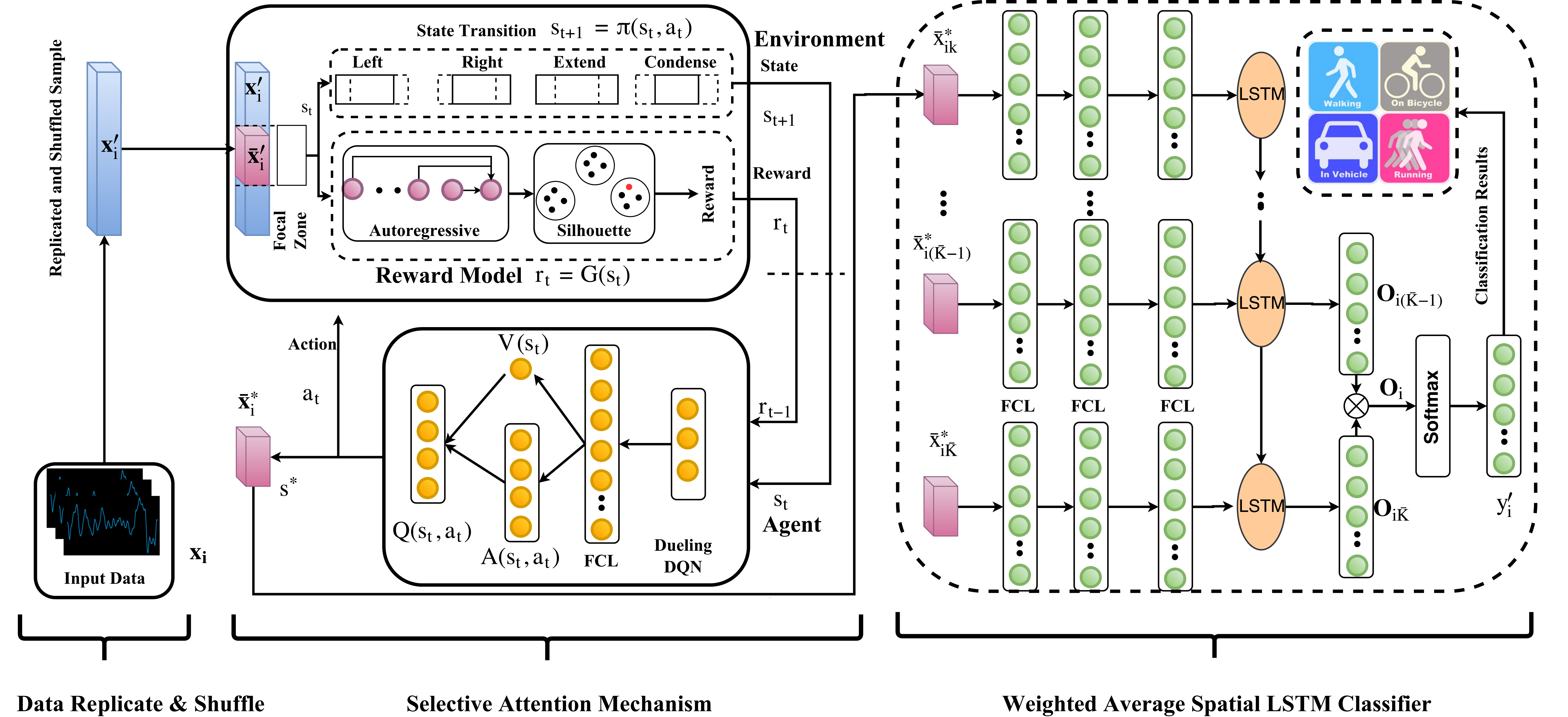}
  \caption{Flowchart of the proposed framework.
   The focal zone $\mathbf{\bar{x}}_i$ is a selected fragment from $\mathbf{x}'_i$ to feed in the state transition and the reward model.
   In each step $t$, one action is selected by the state transition to update $s_t$ based on the agent's feedback. The reward model evaluates the quality of the focal zone to the reward $r_t$.
   The dueling DQN is employed to find the optimal focal zone $\mathbf{\bar{x}}^*_i$ which will be feed into the LSTM based classifier to explore the inter-dimension dependency and predict the sample's label $y'_i$. $FCL$ denotes Fully Connected Layer. The State Transition contains four actions: left shifting, right shifting, extend, and condense. The dashed line indicates the focal zone before the action while the solid line indicates the position of the focal zone after the action.}
  \label{fig:workflow}
  \vspace{-3mm}
\end{figure*}

\section{The Proposed Framework} % (fold)
\label{sec:cognition_detection_framework}
In this section, we present the cognition detection framework in detail. The subjects' brain activity can be measured by a number of methods like EEG, fMRI, MEG. In this paper, we 
%take
exploit EEG 
due to 
%as an example for the 
its unique features such as low-cost, low-energy, privacy, and 
portability.
%, etc.
The proposed framework is 
%shown 
depicted in Figure~\ref{fig:workflow}. 
The main focus of the 
%michael: mentioning algorithm is a bit abrupt. suggest to use approach
%algorithm 
approach is to exploit the latent dependency among different signal dimensions. To this end, the proposed 
%approach 
framework 
contains several components: 1) the replicate and shuffle processing; 2) the selective attention learning; 3) the sequential LSTM-based classification. 
In the following, we will first discuss the motivations of the proposed method and then introduce the aforementioned components in details.

\subsection{Motivation} % (fold)
\label{sub:motivation}
% subsection motivation (end)
How to exploit the latent relationship between EEG signal dimensions is the main focus of the proposed approach. The signals belonging to different cognitions are supposed to have different inter-dimension dependent relationships which contain rich and discriminative information. This information is critical to improve the distinctive signal pattern discovery.

In practice, the EEG signal is often arranged as 1-D vector, the signal is less informative for the limited and fixed element arrangement. The elements order and the number of elements in each signal vector can affect the element dependency.
For example, the inter-dimension dependency in \{0,1,2,3,4\} and \{1,2,3,4,0\} are not reciprocal; similarly, \{0,1,2,3,4\} and \{0,1,1,2,3,4\} are not reciprocal.
In many real-world scenarios, the EEG data are concatenated following the distribution of biomedical EEG channels.
Unfortunately,  the practical channel sequence, with the fixed order and number, may not be suitable for inter-dimension dependency analysis.
 % Meanwhile, the optimal dimension sequence \cite{tan2015lstm} varies with the sensor types and combinations. 
Therefore, we propose the following three techniques to amend the drawback.

First, we replicate and shuffle the input EEG signal vector on dimension-wise in order to provide as much latent dependency as possible among feature dimensions (Section~\ref{sub:repeat_and_shuffle}). 

Second, we introduce a focal zone as a Selective Attention Mechanism (SAM), where the optimal inter-dimension dependency for each sample only depends on a small subset of features. Here, the focal zone is optimized by deep reinforcement learning which has been shown to achieve both good performance and stability in policy learning (Section~\ref{sub:attention_pattern_learning}).

Third, we propose the WAS-LSTM classifier by extracting the distinctive inter-dimension dependency (Section~\ref{sub:classification}).

\subsection{Data Replicate and Shuffle} % (fold)
\label{sub:repeat_and_shuffle}
 
Suppose the input EEG data can be denoted by $\mathbf{X}=\{(\mathbf{x}_i, y_i), i=1,2,\cdots I\}$ where $(\mathbf{x}_i, y_i)$ denotes the 1-D EEG signal, called one \textit{sample} in this paper, and $I$ denotes the number of samples. In each sample, the feature $\mathbf{x}_i\in \mathbb{R}^K$ contains $K$ elements and the corresponding ground truth $y_i\in \mathbb{R}$ is an integer 
that 
denotes the sample's category. Different categories 
%corresponding 
correspond 
to various brain activities. $\mathbf{x}_i$ can be described as a vector with $K$ elements, $\mathbf{x}_i=\{x_{ik},k=1,2,\cdots, K\}$.

To provide more potential inter-dimension spatial dependencies, we propose a method called {\em Replicate and Shuffle} (RS). RS is a two-step feature transformation method which 
%michael: pay attention to English!
%mapping 
maps $\mathbf{x}_i$ to a higher dimensional space $\mathbf{x}'_i$ with more complete element combinations:
$$\mathbf{x}_i\in \mathbb{R}^K \rightarrow \mathbf{x}'_i \in \mathbb{R}^{K'}, K'>K$$
In the first step (Replicate), 
we
replicate $\mathbf{x}_i$ for $h = K'\%K+1$ times where $\%$ denotes remainder operation. Then we get a new vector with length as $h*K$ which is not less than $K'$; in the second step (Shuffle), we randomly shuffle the replicated vector in the first step and intercept the first $K'$ element to generate $\mathbf{x}'_i$. Theoretically, compared to $\mathbf{x}_i$, the number and order of elements in $\mathbf{x}'_i$ are more diverse.
 % and complete inter-dimension dependencies. 
 For instance, set $\mathbf{x}_i = \{1,3,4,2\}$, in which the four elements are arranged in a fixed order and limited combinations, it is difficult to mine the latent pattern in $\mathbf{x}_i$; however, set the replicated and shuffled signal as $\mathbf{x}'_i = \{3,1,2,3,4,4,1,2\}$, the equal difference characteristic is easy to be found in the fragment $\{1, 2, 3,4\}$ (the 2-nd to 5-th elements of $\mathbf{x}'_i$). Therefore, a major challenge in this work is to discover the fragment with rich distinguishable information. To solve this problem, we propose a attention based selective mechanism which is detailed introduced in Section~\ref{sub:attention_pattern_learning}.

 \subsection{Selective Attention Mechanism} % (fold)
 \label{sub:attention_pattern_learning}
In the next process, we attempt to find the optimal dependency which includes the most distinctive information. But $K'$, the length of $\mathbf{x}'_i$, is too large and is computationally expensive. To balance the length and the information content, we introduce the {\em attention mechanism} \cite{cavanagh1992attention} to emphasize the informative fragment in $\mathbf{x}'_i$ and denote the fragment by $\mathbf{\bar{x}}_i$, which is called \textit{focal zone}.
Suppose $\mathbf{\bar{x}}_i \in \mathbb{R}^{\bar{K}}$ and $\bar{K}$ denotes the length of the focal zone. For 
%simplification, 
simplicity, 
we continue 
to 
denote the $k$-th element by $\mathbf{\bar{x}}_{ik}$ in the focal zone. To optimize the focal zone, we employ deep reinforcement learning as the optimization framework for its excellent performance in policy optimization \cite{mnih2015human}.

%michael: to be contiued.
% subsection overview (end)
\vspace{2mm}
\noindent{\bf Overview.} 
As shown in Figure~\ref{fig:workflow}, the focal zone optimization includes two key components: the {\em environment} (including state transition and reward model), and the {\em agent}. Three elements (the state $s$, the action $a$, and the reward $r$) are exchanged in the interaction between the environment and the agent. In the following we elaborate 
%on 
these three elements which are crucial to our proposed deep reinforcement learning model:
\begin{itemize}
    \item The \textbf{state} $\mathcal{S}=\{s_t, t=0,1,\cdots,T\}\in \mathbb{R}^2$ describes the position of the focal zone, where $t$ denotes the time stamp. In the training, $s_0$ is initialized as $s_0=[(K'-\bar{K})/2, (K'+\bar{K})/2]$. Since the focal zone is a shifting fragment on 1-D $\mathbf{x}'_i$, we design two parameters to define the state: $s_t = \{start^t_{idx},end^t_{idx}\}$, where $start^t_{idx}$ and $end^t_{idx}$ separately denote the start index and the end index of the focal zone\footnote{E.g., for a random $\mathbf{x}'_i = [3,5,8,9,2,1,6,0]$, the state $\{start^t_{idx}=2, end^t_{idx}=5\}$ is sufficient to determine the focal zone as $[8,9,2,1]$.}.
    \item The \textbf{action} $\mathcal{A}=\{a_t,t=0,1,\cdots,T\}\in \mathbb{R}^4$ describes which the agent could choose to act on the environment. In our case, we define 4 categories of actions for the focal zone (as described in the \textbf{State Transition} part in Figure~\ref{fig:workflow}): left shifting, right shifting, extend, and condense. Here at time stamp $t$, the state transition only choose one action to implement following the agent's policy $\pi$: $s_{t+1}=\pi(s_t,a_t)$.
    \item The \textbf{reward} $\mathcal{R}=\{r_t,t=0,1,\cdots,T\}\in \mathbb{R}$ is calculated by the reward model, which will be detailed later. The reward model $\Phi$: $r_{t}=\Phi(s_t)$ receives the current state and returns an evaluation as the reward.
    % The reward model will be detailed in Section~\ref{sub:reward_model}.
    \item We employ the Dueling DQN (Deep Q Networks \cite{wang2015dueling}) as the optimization \textbf{policy} $\pi(s_t,a_t)$, which is enabled to learn the state-value function efficiently. Dueling DQN learns the Q value $V(s_t)$ and the advantage function $A(s_t,a_t)$ and combines them: $Q(s_t, a_t)\leftarrow V(s_t), A(s_t,a_t)$.
    The primary reason we employ a dueling DQN to optimize the focal zone is that it updates all the four Q values
    % \footnote{Since we have four actions in $a_t$, the $Q(s_t, a_t)$ contains 4 Q values. The arrangement is similar to the one-hot label.} 
    at every step while other policy only updates one Q value at each step.
\end{itemize}

\noindent\textbf{Reward Model.} Next, we 
%detailedly 
introduce the design of the reward model, 
which
% for it 
is one 
%crucial 
important contribution of this paper. The purpose of 
the 
reward model is to evaluate how the current state impacts our final target which refers to the classification performance in our case. Intuitively, the state which can lead to the better classification performance should have a higher reward: $r_t=\mathcal{F}(s_t)$. As a result, in the standard reinforcement learning framework, the original reward model regards the classification accuracy as the reward. $\mathcal{F}$ refers to the WAS-LSTM.
%, the WAS-LSTM works on EEG data which is collected at one sampling point, but the normal LSTM works on the temporal data which is a stack of a number of 1-D EEG data. On the other words, every single sample of the normal LSTM is 2-D data.
Note, WAS-LSTM focuses on the spatial dependency between different dimensions at the same time-point while the normal LSTM focuses on the temporal dependency between a sequence of samples collected at different time-points.
% In the standard reinforcement learning framework, intuitively, the reward model, $r_t=\mathcal{F}(s_t)$, is directly determined by the classifier. In our case, $\mathcal{F}$ refers to the Deep Learning based Classifier (DLC).
However, 
%the 
WAS-LSTM requires considerable training time, which will dramatically increase the optimization time of the whole algorithm. In this section, we propose an alternative method to calculate the reward: construct a new reward function $r_t=\mathcal{G}(s_t)$ which is positively related with $r_t=\mathcal{F}(s_t)$. Therefore, we can employ $\mathcal{G}$ to replace $\mathcal{F}$. Then, the task is changed to construct a suitable $\mathcal{G}$ which can evaluate the inter-dimension dependency in the current state $s_t$ and feedback the corresponding reward $r_t$. We propose an alternative $\mathcal{G}$ composed by three components: the autoregressive model \cite{akaike1969fitting} to exploit the inter-dimension dependency in $\mathbf{x}'_i$, the Silhouette Score \cite{laurentini1994visual} to evaluate the similarity of the autoregressive coefficients, and the reward function based on the silhouette score.

The autoregressive model \cite{akaike1969fitting} receives the focal zone $\mathbf{\bar{x}}_i$ and specifies that how the last variable depends on its own previous values.
% $$\bar{x}_{i\bar{K}} = \sum_{j=1}^{p}\varphi_j \bar{x}_{i(\bar{K}-j)}+ C +\bar{\varepsilon}$$
% where $p$ is the order of the autoregressive model, $C$ indicates a constant, and $\bar{\varepsilon}$ indicates the white noise. 
% From this equation, we can infer that
% the autoregressive coefficient $\boldsymbol{\varphi}=\{\varphi_j,j=1,2,\cdots,p\} \in \mathbb{R}^p$ incorporates the dependent relationship in the focal zone. 
Then, to evaluate how rich information is taken in the autoregressive coefficients, we employ silhouette score \cite{lovmar2005silhouette} $ss_t$
%\footnote{\url{https://en.wikipedia.org/wiki/Silhouette_(clustering)}}
to interpret the consistence of $\boldsymbol{\varphi}$. 
The silhouette score measures how similar an object is to its own cluster compared to other clusters and a high silhouette value indicates that the object is well matched to its own cluster and poorly matched to neighboring clusters. Specifically, in our case, the higher silhouette score means that $\boldsymbol{\varphi}$ can be better clustered and the focal zone $\mathbf{\bar{x}}_i$ is be easier classified.
At last, based on the $ss_t\in[-1,1]$, we design a \textbf{reward function}:
$$r_t = \frac{e^{ss_t+1}}{e^2-1}-\beta \frac{\bar{K}}{K'}$$
The function contains two parts, the first part is a normalized exponential function with the exponent $ss_t+1 \in[0,1]$, 
%this part 
which encourages the reinforcement learning algorithm to search the better $s_t$ w that leads to a higher $ss_t$. 
The motivation of the exponential function is that: the reward growth rate is increasing with the silhouette score's increase\footnote{For example, for the same silhouette score increment 0.1, $ss_t: 0.9\rightarrow 1.0$ can earn higher reward increment than $ss_t: 0.1\rightarrow 0.2$.}. The second part is a penalty factor for the focal zone length to keep the bar shorter and the $\beta$ is the penalty coefficient.

In summary, the aim of focal zone optimization is to learn
% the optimal state $s*$ and
the optimal focal zone $\mathbf{\bar{x}}^*_i$ which can lead to the maximum reward. The optimization totally iterates $N=n_e*n_s$ times where $n_e$ and $n_s$ separately denote the number of episodes and steps \cite{wang2015dueling}. $\varepsilon$-greedy method \cite{tokic2010adaptive} is employed in the state transition.

\subsection{Weighted Average Spatial LSTM Classifier} % (fold)
\label{sub:classification}
% The EEG data classification section
In this section, we propose Weighted Average Spatial LSTM classification for two purposes. The first attempt is to capture the cross-relationship among feature dimensions in the optimized focal zone $\mathbf{\bar{x}}^*_i$. The LSTM-based classifier is widely used for its excellent sequential information extraction ability which is approved in several research areas such as natural language processing \cite{gers2001lstm}. Compared to other commonly employed spatial feature extraction methods, such as Convolutional Neural Networks, LSTM less dependent on the hyper-parameters setting. However, the traditional LSTM focuses on the temporal dependency among a sequence of samples. Technically, the input data of traditional LSTM is 3-D tensor shaped as $[n_b, n_t, \bar{K}]$ where $n_b$ and $n_t$ denote the batch size and the number of temporal samples, separately. 

In this paper, we transpose the input data as $[n_b, n_t, \bar{K}]\rightarrow[n_b, \bar{K}, n_t]$ following the equation $(A^T)_{ijk} = A_{ikj}$, in which form, each sample has shape $[\bar{K}, n_t]$ and the WAS-LSTM pays attention to each sample column and explores the latent dependencies between the various elements in the same column. WAS-LSTM aims to capture the dependency among various dimensions at one temporal point, therefore, we set $n_t=1$.

The second advantage of WAS-LSTM is that it could stabilize the performance of LSTM via moving average method.
 % \cite{lipton2015learning}.
In LSTM, each cell's output contains the information before it, however, the neural network’s convergence and stability are fluctuated over different times of training. To enhance the convergence and stability, we calculate the LSTM outputs $\mathbf{O}_i$ by averaging the weighted past two outputs instead of only the final one (Figure~\ref{fig:workflow}):
$$\mathbf{O}_i = w_1\mathbf{O}_{i(\bar{K}-1)}+w_2\mathbf{O}_{i\bar{K}}$$
where $w_1$ and $w_2$ are the corresponding weights which can adjust the importance proportion of $\mathbf{O}_{i(\bar{K}-1)}$ and $\mathbf{O}_{i\bar{K}}$. The weights can be automatically learned by the neural network \cite{zhang2017mindid} or be manually set. In this paper, we simply manually set $w_1=w_2=0.5$ in order to save computing resources.
The predicted label is calculated by $y'_i = \mathcal{L}(\mathbf{\bar{x}}^*_i)$ where $\mathcal{L}$ denotes the LSTM algorithm. $\ell_2$-norm (with parameter $\lambda$) is adopted as regularization to prevent overfitting. The sigmoid activation function is used on hidden layers. The loss function is cross-entropy and is optimized by the AdamOptimizer algorithm.

\begin{figure}[!t]
\centering
\subfloat[]{\includegraphics[width=0.42\linewidth]{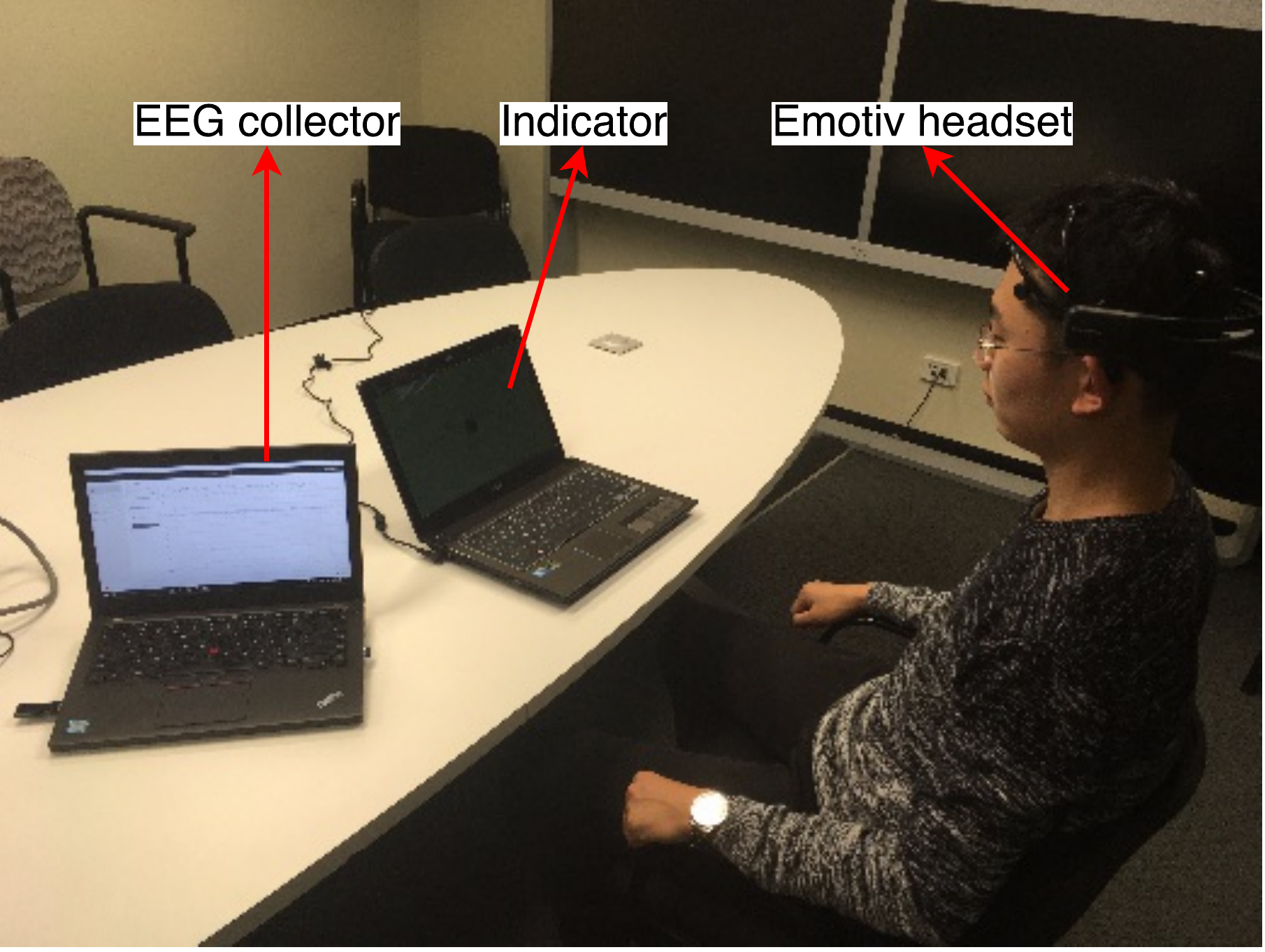}%
\label{fig_collection}}
\hfil
\subfloat[]{\includegraphics[width=0.56\linewidth]{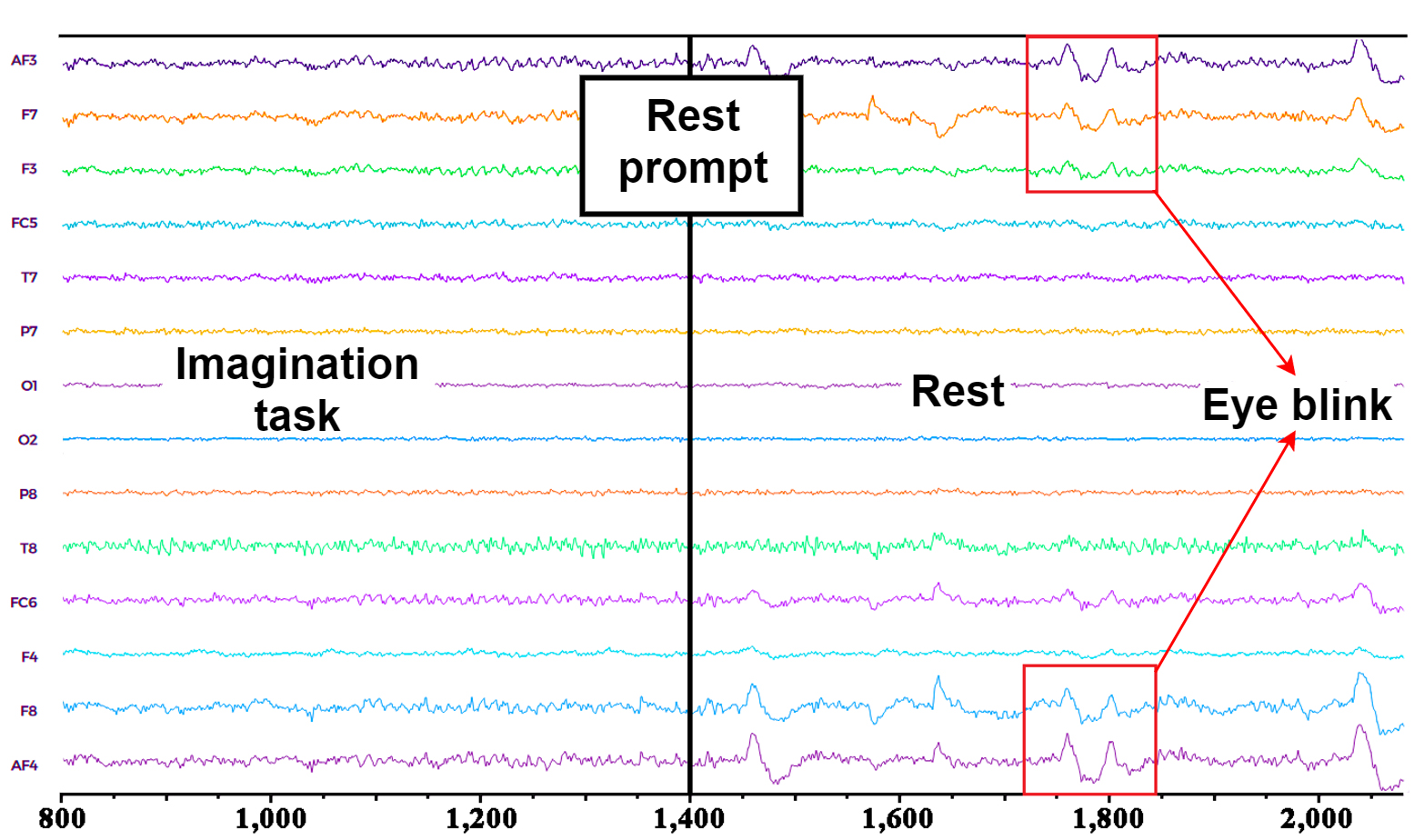}%
\label{fig_eegsignal}}
\caption{EEG data collection: (a) collection scenario and (b) the gathered EEG signal}
\label{fig_data_collection}
\end{figure}

\section{Experiments} % (fold)
\label{sec:experiments}
In this section, we design local real-world experiments to evaluate the efficiency and effectiveness of the proposed framework. First, the experimental setting is reported. Then, we compare our model with competitive state-of-the-art baselines and evaluate the performance in detail.
% at last, 
Finally, we investigate the impact of crucial factors such as the framework latency and the reward model.

% To examine the adaptability and consistency of our model, we further evaluate our proposed model on a limited but easy-to-deploy dataset.

\begin{figure*}[!t]
\centering
\subfloat[]{\includegraphics[width=1.8in]{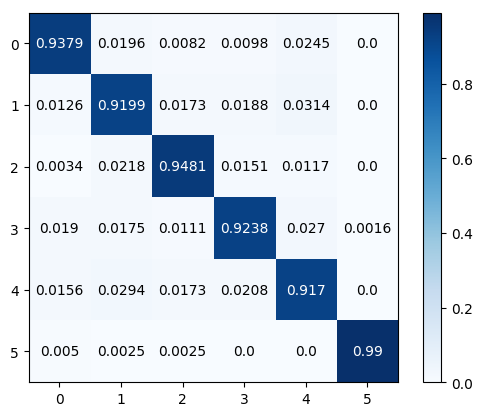}%
\label{fig_ccm}}
\hfil
\subfloat[]{\includegraphics[width=2in]{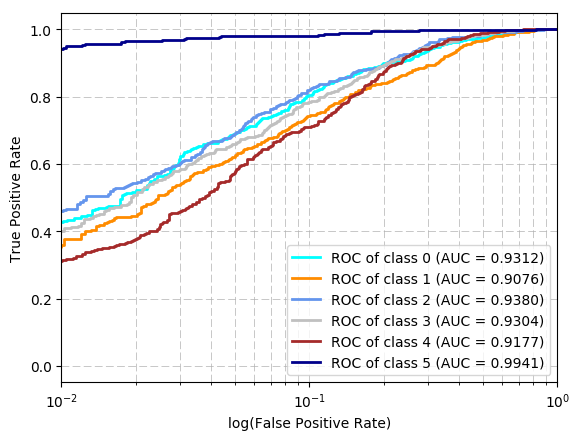}%
\label{fig_roc}}
\hfil
\subfloat[]{\includegraphics[width=2.7in]{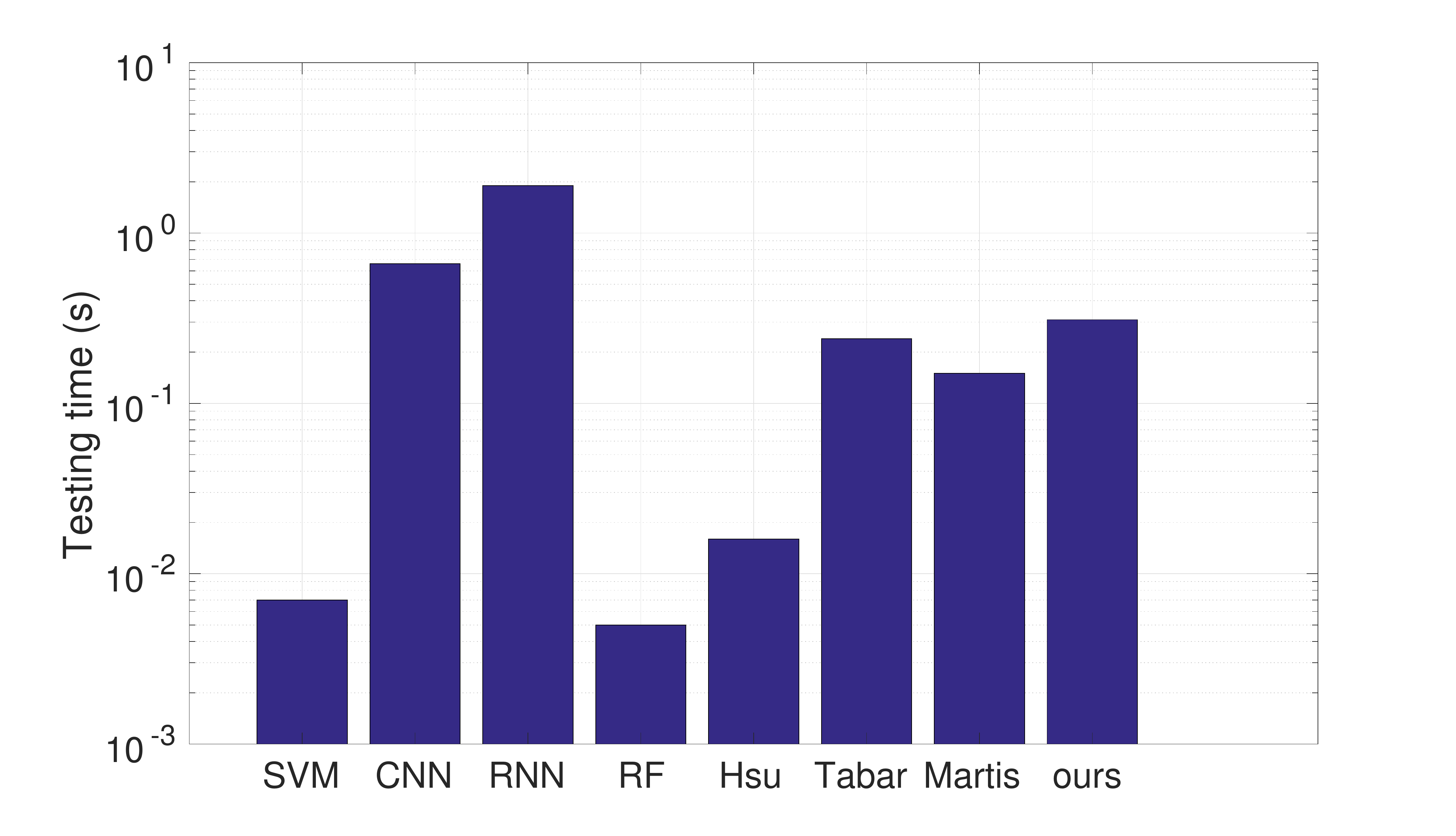}%
\label{fig_latency}}
\caption{Recognition results: (a) confusion matrix, (b) ROC curves with AUC scores, and (c) latency}
\label{fig_cm_roc}
\end{figure*}

\begin{table}[!t]
\renewcommand{\arraystretch}{1.3}
\caption{\MakeUppercase{Imagery Action, label, and corresponding commands in case studies}}
\label{tab_command}
\centering
\begin{tabular}{llll}
\hline
\textbf{Imagery Action} & \textbf{Label} & \textbf{Typing Commands} & \textbf{Robot Commands} \\ \hline
Upward & 0 & Up & Forward \\
Downward & 1 & Cancel & Turn Left \\
Leftward & 2 & Left & Grasp \\
Rightward & 3 & Right & Loose \\
Middle Cycle & 4 & Nothing & Nothing \\
Eye-closed & 5 & Confirm & Stop/Start \\ \hline
\end{tabular}
\end{table}

\subsection{Experimental Setting} % (fold)
\label{sub:experimental_setting}
We conduct the EEG collection by using a portable and easy-to-use commercialized Emotiv Epoc+ headset. The headset contains 14 channels and the sampling rate is 128 Hz. The local dataset can be accessed from this link\footnote{\url{https://drive.google.com/open?id=0B9MuJb6Xx2PIM0otakxuVHpkWkk}}. This experiment is carried out using 7 subjects (4 males and 3 females) aged from 23 to 26. During the experiment, the subject wearing the \textit{Emotiv Epoc+}\footnote{\url{https://www.emotiv.com/product/emotiv-epoc-14-channel-mobile-eeg/}} EEG collection headset, faces the computer screen and focuses on the corresponding \textit{hint} which appears on the screen (shown in Figure~\ref{fig_collection}). EEG signals are recorded when the subject is imaging certain actions (without any physical action). The certain actions contains: upward arrow, downward arrow, leftward arrow, rightward arrow, and a cycle. Beyond that, the EEG signals that the subject stays relaxation with eye closed are also recorded. In total, there are 6 categories of EEG signals. The imagery action 
%which 
associated with brain activities and the corresponding labels used in this paper are listed in Table~\ref{tab_command}. In summary, this experiment contains 241,920 samples with 34,560 samples for each subject. For each participant, the dataset is divided into a training set and a testing set. The training set contains 31,104 samples and the testing set contains 3,456 samples. The classification results are evaluated by a number of metrics including accuracy, precision, recall, F-1 score, confusion matrix, ROC (Receiver Operating Characteristic) curve, AUC (Area Under Curve) score. 
% In order to distinguish with the aforementioned \textit{eegmmidb} dataset, we name this dataset as \textit{emotiv}. 

\subsection{Overall Comparison and Analysis} % (fold)
\label{sub:results_and_analysis}
In the training stage, based on the tuning experience, the hyper-parameters setting are listed as follows.
In the selective attention learning: the order of autoregressive is 3; $\bar{K}=128$, the Dueling DQN has 4 
%lays 
layers
and the node number in each layer are: 2 (input layer), 32 (FCL), 4 ($A(s_t,a_t)$) + 1 ($V(s_t)$), 4 (output). The decay parameter $\gamma =0.8$, $n_e=n_s=50$, $N=2,500$, $\epsilon=0.2$, learning rate$ =0.01$, memory size $ =2000$, length penalty coefficient $\beta=0.1$, and the minimum length of focal zone is set as 10. In the deep learning classifier: the node number in the input layer equals to the number of feature dimensions, three hidden layers with 164 nodes, two layers of LSTM cells (164 cells) and one output layer (6 nodes). The learning rate $ =0.001$, $\ell_2$-norm coefficient $\lambda=0.001$, forget bias $=0.3$, batch size $ =9$, and iterate for 1000 iterations.

To demonstrate the efficiency of our approach, we compare our model with several competitive state-of-the-art methods:
\begin{itemize}
  \item \textbf{Hsu} \cite{hsu2015assembling} extracts several potential features, including amplitude modulation, spectral power and asymmetry ratio, adaptive autoregressive model, and wavelet fuzzy approximate entropy (wfApEn), followed by a SVM classifier, to classify the binary motor imagery EEG signals.
  \item \textbf{Tabar et al.} \cite{tabar2016novel} combine convolutional neural networks (CNN) and stacked Autoencoder (SAE) to automatically classify EEG data.
  \item \textbf{Martis et al.} \cite{martis2015epileptic} artificially extract several nonlinear features on different EEG frequency bands (including delta, theta, lower alpha, upper alpha, lower beta, upper beta and lower gamma) and forward to SVM with radial basis function kernel.
\end{itemize}

Table~\ref{tab_comparison} shows the overall comparison between our approach with non-DL baselines, DL baselines, and the state-of-the-art models. RF denotes Random Forest, AdaB denotes Adaptive Boosting, LDA denotes Linear Discriminant Analysis. In addition, the key parameters of the baselines are listed here: Linear SVM ($C=1$), RF ($n=200$), KNN ($k=3$). In LSTM, $n_{steps}=5$, another set is the same as the WAS-LSTM classifier, along with the GRU (Gated Recurrent Unit). The CNN contains 2 stacked convolutional layers (both with stride $[1,1]$, patch $[2,2]$, zero-padding, and the depth are 4 and 8, separately.), one pooling layer (stride $[1,2]$, zero-padding), and one fully connected layer (164 nodes). Relu activation function is employed in the CNN.

\begin{table}[]
\centering
\caption{\MakeUppercase{overall comparison with the state-of-the-art baselines. DL denotes deep learning.}}
\label{tab_comparison}
\resizebox{\linewidth}{!}{
\begin{tabular}{llllll}
\hline
\multirow{2}{*}{\textbf{Baselines}} & \multirow{2}{*}{\textbf{Methods}} & \multicolumn{4}{c}{\textbf{Metrics}} \\ \cline{3-6} 
 &  & \textbf{Acc} & \textbf{Pre} & \textbf{Rec} & \textbf{F1-score} \\ \hline
\multirow{5}{*}{\textbf{Non-DL}} & \textbf{SVM} & 0.2569 & 0.2737 & 0.2569 & 0.2577 \\
 & \textbf{RF} & 0.8041 & 0.8071 & 0.8041 & 0.8048 \\
 & \textbf{KNN} & 0.8539 & 0.8563 & 0.8539 & 0.8544 \\
 & \textbf{AB} & 0.2506 & 0.2039 & 0.2506 & 0.1557 \\
 & \textbf{LDA} & 0.2595 & 0.2761 & 0.2595 & 0.2618 \\ \hline
\multirow{3}{*}{\textbf{DL}} & \textbf{LSTM} & 0.2609 & 0.2447 & 0.2348 & 0.2354 \\
 & \textbf{GRU} & 0.2521 & 0.271 & 0.2696 & 0.2701 \\
 & \textbf{CNN} & 0.725 & 0.724 & 0.7237 & 0.7238 \\ \hline
\multirow{3}{*}{\textbf{\begin{tabular}[c]{@{}l@{}}The state-\\ of-the-art\end{tabular}}} & \textbf{\cite{hsu2015assembling}} & 0.8965 & 0.9011 & 0.8926 & 0.8968 \\
 & \textbf{\cite{tabar2016novel}} & 0.7894 & 0.7938 & 0.8013 & 0.7975 \\
 & \textbf{\cite{martis2015epileptic}} & 0.8891 & 0.8932 & 0.8765 & 0.8848 \\ \hline
\textbf{} & \textbf{WAS-LSTM} & 0.9026 & 0.9125 & 0.9003 & 0.9064 \\
\textbf{} & \textbf{SAM+WAS-GRU} & 0.9135 & 0.9188 & 0.9395 & 0.9378 \\
\textbf{} & \textbf{Ours} & \textbf{0.9363} & \textbf{0.9394} & \textbf{0.9398} & \textbf{0.9396} \\ \hline
\end{tabular}
}
\vspace{-3mm}
\end{table}

The observations in Table~\ref{tab_comparison} 
%illustrate 
show that our approach outperforms all the baselines by achieving the highest accuracy of \textbf{0.9363} on the 6-class classification. In addition, our model (SAM +WAS-LSTM) performs better than the solo WAS-LSTM, which demonstrates that the selective attention mechanism has a positive contribution to the classification. The confusion matrix, ROC curves, and  AUC scores of the proposed framework are reported in Figure~\ref{fig_cm_roc}. We can observe that the last class, representing the eye-closed state, obtains the best performance compared to other 5 classes. This demonstrates that the eye-closed state is 
the easiest to be recognized, which is reasonable while all the other classes are in eye-open state and are easier to be interrupted by the environmental factors. 
Moreover, through the results comparison of SAM+WAS\-GRU and our model (SAM+WAS\-LSTM) (Table~\ref{tab_comparison}), we can observe that the latter achieves higher performance which indicates (0.9363 $>$ 0.9135) the LSTM slightly outperforms GRU in our scenarios. The reason can be inferred is that LSTM can remember longer sequences than GRU.

\subsection{Impact of Key Factor} % (fold)
\label{sub:impact_of_key_factor}

\subsubsection{Latency} % (fold)
\label{sub:latency}

% subsection latency (end)
To design effective and real-world cognitive interactive applications, both the accuracy and latency of intent recognition are equally important. Subsequently, we compare the latency of the proposed framework with several typical state-of-the-art algorithms and the results
are presented in Figure~\ref{fig_latency}. It is observed that our approach has competitive latency compared with other methods. The overall latency
is less than 1 second. The deep learning based techniques in this
work do not explicitly lead to extra latency.

\subsubsection{Reward Model} % (fold)
\label{sub:reward_model}

\begin{figure}[t]
\centering
\begin{minipage}[b]{0.44\linewidth}
\centering
\includegraphics[width=\textwidth]{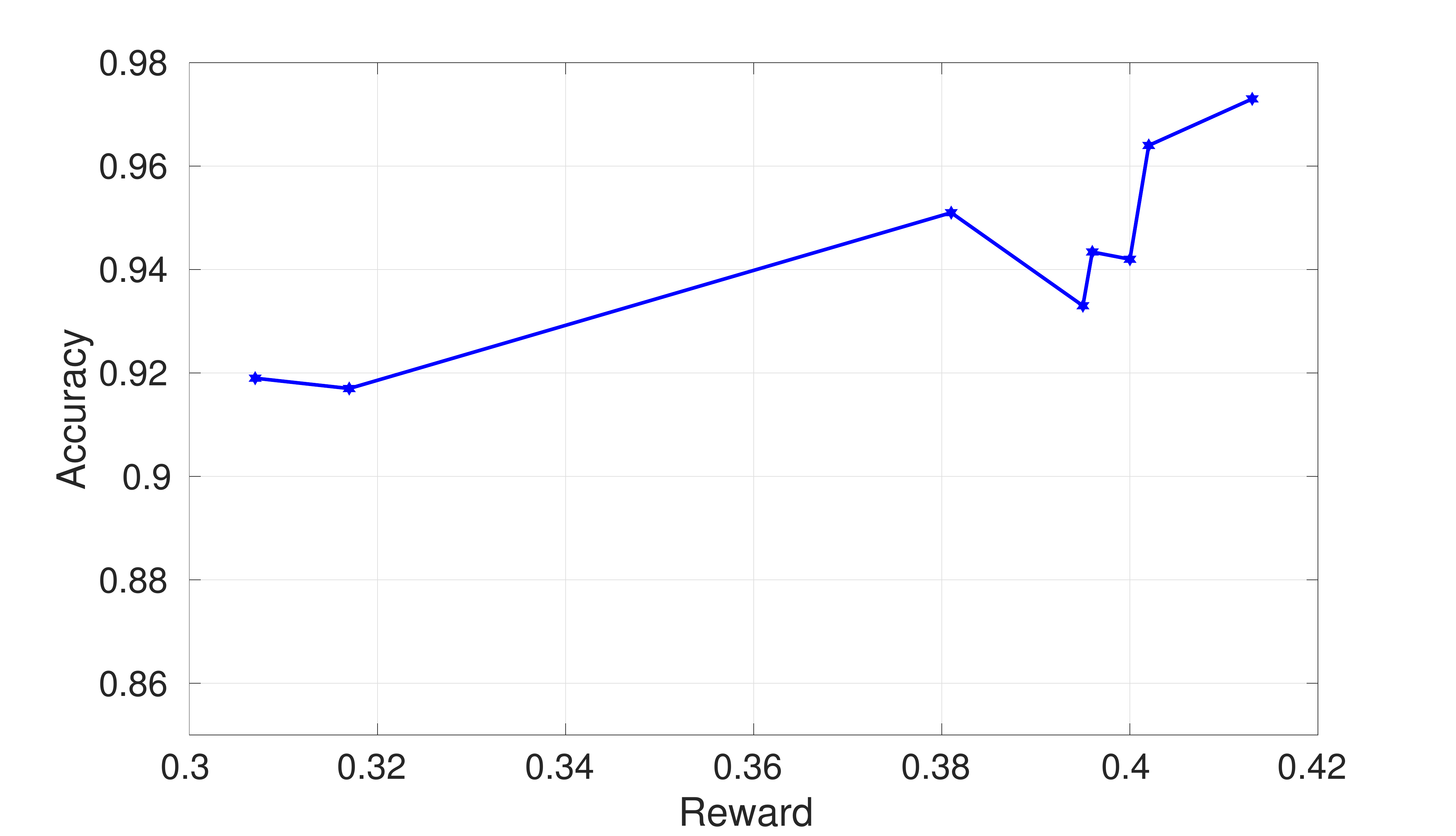}
\caption{The relationship between the classifier accuracy and the reward. The correlationship coefficient is 0.8258 while the p-value is 0.0115.}
\label{fig:reward_acc}
\end{minipage}
\hspace{1mm}
\begin{minipage}[b]{0.44\linewidth}
\centering
\includegraphics[width=\textwidth]{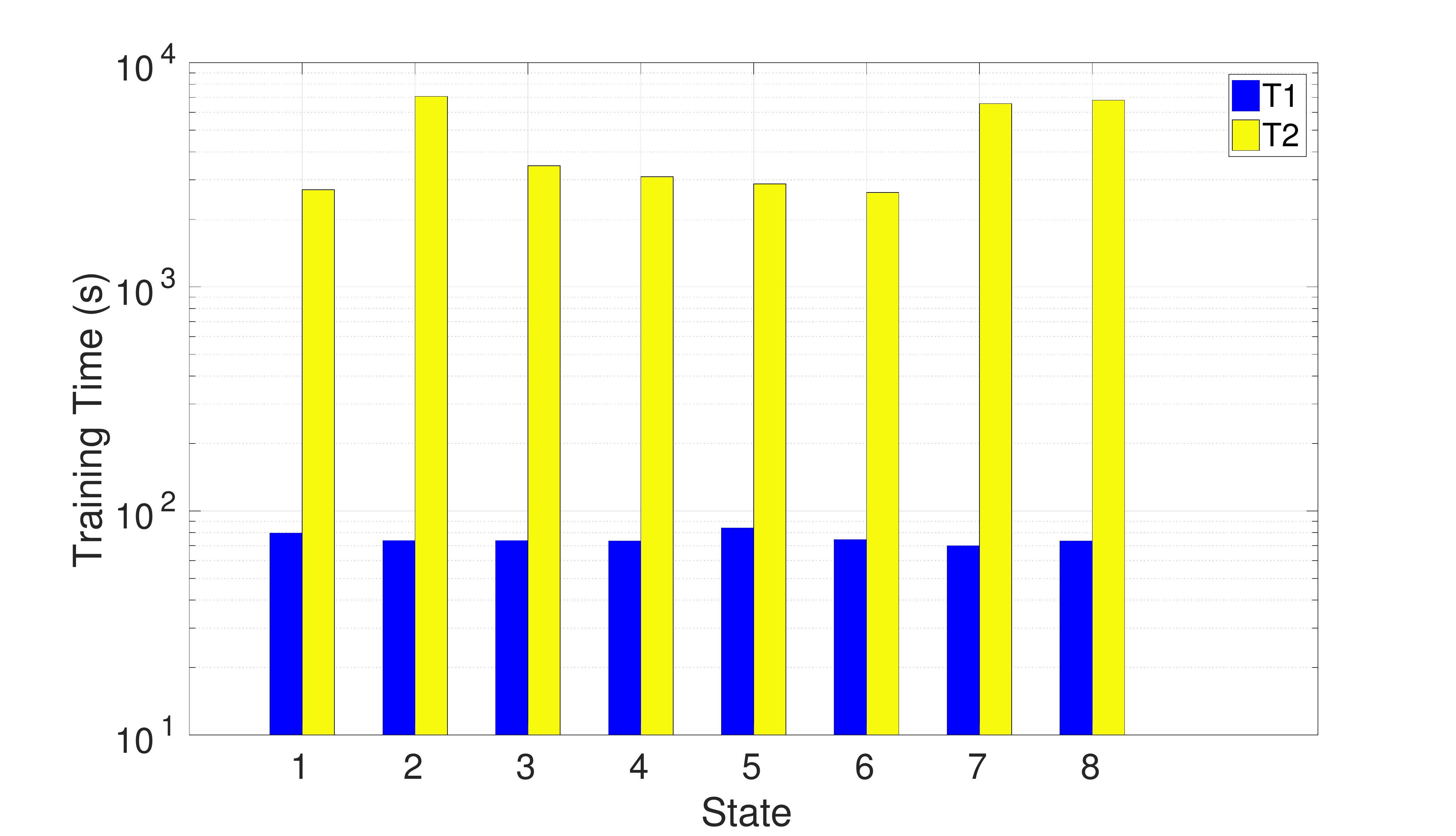}
\caption{Reward model training time in various states. T1 and T2 separately denote the training time in reward model $\mathcal{G}$ and $\mathcal{F}$. }
\label{fig:time_compare}
\end{minipage}
\vspace{-3mm}
\end{figure}

% subsection reward_model (end)
Furthermore, we conduct extensive experiments to demonstrate the efficiency of the proposed reward model $\mathcal{G}$.
First, we measure a batch of data pairs of the reward (represents the reward of $\mathcal{G}$) and the WAS-LSTM classifier accuracy (represents the reward of $\mathcal{F}$). The relationship between the reward and the accuracy is shown in Figure~\ref{fig:reward_acc}. The figure illustrates that the accuracy has an approximately linear relationship with the reward. The correlations coefficient is 0.8258 (with p-value as 0.0115), which 
%demonstrates 
shows that the accuracy and reward are highly positive related. As a result, we can estimate $\mathop{\arg\max}\limits_{\mathbf{\bar{x}^*}}\mathcal{F}$ by $\mathop{\arg\max}\limits_{\mathbf{\bar{x}^*}}\mathcal{G}$. Moreover, another experiment is carried 
%on 
out to measure the single step training time of two reward models $\mathcal{G}$ and $\mathcal{F}$. The training times are marked as T1 and T2, respectively. Figure~\ref{fig:time_compare} qualitatively shows that T2 is much higher than T1 (8 states represent 8 different focal zones). Quantitatively, the sum of T1 over 8 states is $35,237.41$ seconds while the sum of T2 is $601.58$ seconds. These results demonstrate that the proposed approach, designing a $\mathcal{G}$ to approximate and estimate the $\mathcal{F}$, saves $\mathbf{98.3\%}=1-601.58/35237.41$ training time in focal zone optimization.

\section{Case Study} % (fold)
\label{sec:case_study}

\begin{figure*}[!t]
\centering
\subfloat[]{\includegraphics[width=3in]{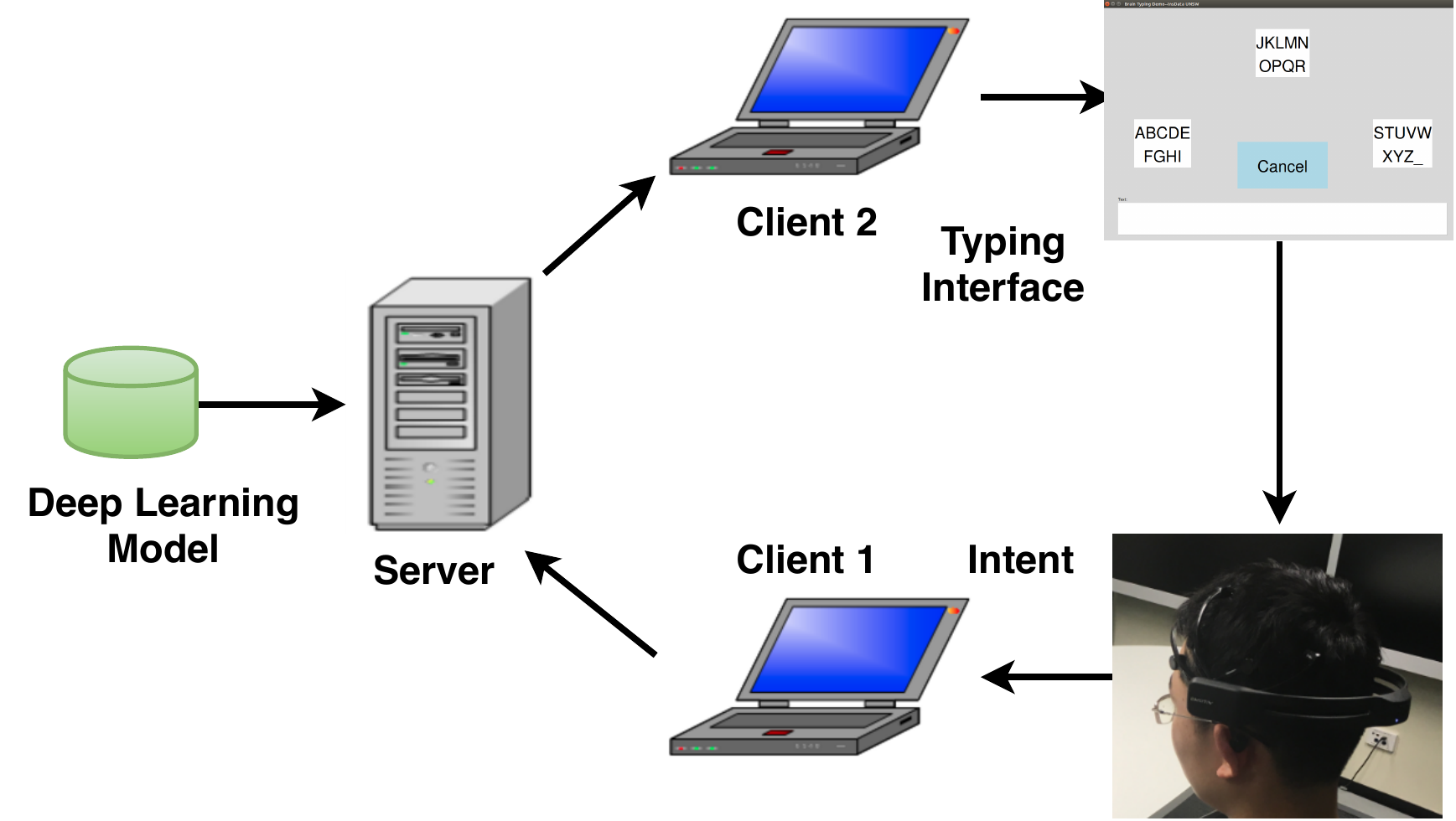}%
\label{fig_BTS}}
\hfil
\subfloat[]{\includegraphics[width=2.4in]{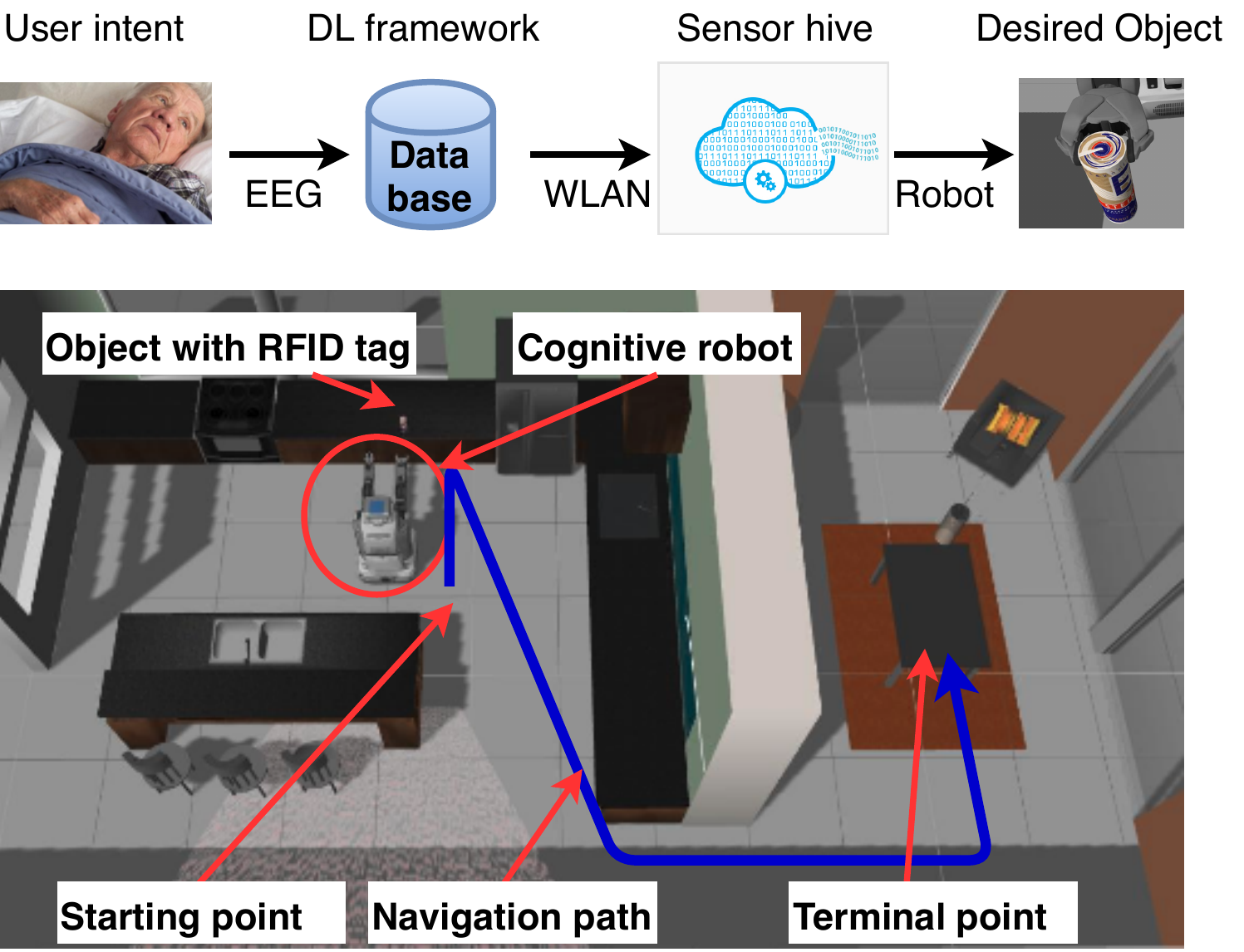}%
\label{fig_robot}}
\caption{Human-thing cognitive applications: (a) brain typing system and (b) cognitive robot in IoT scenario
 % and WLAN denotes Wireless Local Area Network
}
\label{fig_application}
\vspace{-3mm}
\end{figure*}

Inspired by the high accuracy and low latency of our proposed framework for human intent recognition, we proceed to develop two real-world cognitive IoT prototypes, namely, (1) a brain typing system
 % \cite{zhang2017converting} 
 (2) mind-controlled assistive robot for the smart home.
  % \cite{zhang2017intent}.  
%michael: not sure it is good to give two references here. Reviewers may ask what is new in this paper? 

\subsection{Brain Typing System} 
\label{sub:brain_typing_system}
% \textcolor{red}{may cut some content of this section?}
Due to the high intent recognition accuracy, we develop an online brain typing system to convert user's thoughts to texts. 
The video demo clip can be found at the given link\footnote{\url{https://youtu.be/Dc0StUPq61k}}.
The brain typing system (Figure~\ref{fig_BTS}) consists of two components: the {\em pre-trained deep learning model} and the {\em online BCI system}. 
The pre-trained deep learning model, which is trained offline, aims to accurately recognize the user's typing intent in real time.
 % This model is central to the operation of the brain typing system and is the main contribution of this paper.
The online system contains 5 components: the EEG headset, the client 1 (data collector), the server, the client 2 (typing command receiver), and the typing interface. 
The user wears the Emotiv EPOC+ headset
which collects EEG signals and sends
the data to 
%the 
client 1 through a Bluetooth connection. The raw EEG signals are transported to the server through a TCP connection.
% The server 
%receives the user's brainwave signal and feeds the incoming EEG signals to the pre-trained deep learning model. The model produces a classification decision and converts it to the corresponding typing command which is sent to client 2 through a TCP connection. The typing interface receives the command and manifests the appropriate typing action.

Specifically, the typing interface (up right corner in Figure~\ref{fig_BTS}) can be divided into three levels: the initial interface, the sub-interface, and the bottom interface. 
All the interfaces have similar structure: three \textit{character blocks} (separately distributed in left, up, and down directions), a \textit{display block}, and a \textit{cancel button}. The display block shows the typed output and the cancel button is used to cancel the last operation. 
The typing system in total includes $27=3*9$ characters (26 English alphabets and the space bar) and all of them are separated into 3 character blocks (each block contains 9 characters) in the initial interface. 
Overall, there are 3 alternative selections and each selection will lead to a specific sub-interface which contains 9 characters. 
Again, the $9=3*3$ characters are divided into 3 character blocks and each of them is connected to a bottom interface. In the bottom interface, each block represents only one character. 
% As an example, Figure~\ref{fig:typingprocedure} shows the procedure to type the character `I'.

In the brain typing system, there are 5 commands to control the interface: `left', `up', `right', `cancel', and `confirm'.
 Each command corresponds to a specific motor imagery EEG category (as shown in Table~\ref{tab_command}). Since the user can hardly concentrate for a long time (usually, 
 % the human being can only keep high-concentrating state for 
 less than 10 seconds), the brain activity may represent none of the valid commands sometimes. Nevertheless, the proposed deep learning framework cannot distinguish the invalid brain activity, we leave one specific brain category to represent the invalid signal. If the individual's brain signal is not in any of the 5 valid categories, it is classified as the invalid category and the brain typing system will do nothing under this situation\footnote{Similarly, in the cognitive robot case, 
 % has one category corresponding to the invalid command and 
 the robot will remain the previous state under the invalid command.}. 
 Moreover, based on the experiments results in Section~\ref{sub:results_and_analysis}, the eye-closed state has the highest precision and accuracy, therefore, we select this state as the confirmation command for the reason that `confirmation' is the most crucial command in typing system. To type every single character, the interface is supposed to accept 6 commands. Consider typing the letter `I' as an example. The sequence of commands to be entered is as follows:
`left' (choose the left block with characters $A\sim I$), `confirm', `right' (choose the right block with characters $G\sim I$), `confirm', `right' (choose the right block with characters $I$), `confirm'. 
 
In our practical deployment, the sampling rate of Emotiv EPOC+ headset is set as 128Hz, which means the server can receive 128 EEG recordings each second. Since the brainwave signal varies rapidly and is very easy to be affected by noises, the EEG data stream is sent to server {\it 
%each 
every half second}, which means that the server receives 64 EEG samples each time. The 64 EEG samples are classified by the deep learning framework and generate 64 categories of intents. we calculate the mode of 64 intents and regard the mode as the final intent decision. Furthermore, to achieve steadiness and reliability, the server sends the command to client 2 only if \textit{three consecutive decisions remain consistent}. After the command is sent, the command list will be reset and the system will wait until the next three consistent decisions are made.
% \vspace{-2mm} 

\subsection{Cognitive Robot} % (fold)
\label{sub:cognitive_robot}
Another important application for BCI-inspired Internet of Things is extending the orientation of smart homes by integrating the subject's intent and the real-world IoT objects to effectively control things of interest (TOIs). 

To demonstrate the feasibility of the proposed framework, we report the second use case as implementing cognitive interactivity in an IoT-based smart home system. 
The IoT-based smart home is equipped with sensors, wherein IR sensors, ambient sound, heat, as well as contact sensors are mounted on furniture and used in the home environment in a non-intrusive manner. In our case, within the smart home environment which is perceived by the embedded sensor-networks, a simulated robot is cognitively navigated to perform a routine task. 
In the specific scenario, the robot, learns user's intent from EEG recordings via the proposed framework, to take the IoT object (e.g., a can of beverage) from a table in the kitchen and put it in a table in the living room. The desired object is aggregate with RFID tag which helps to identify the location. The IoT scenario is depicted in Figure~\ref{fig_robot} and the demo can be found at here\footnote{\url{https://youtu.be/VZYX1095Vkc}}. 
The user's intent is carried in the EEG recordings which are forwarded to the deep learning based framework for interpretation. The recognized intent is send to sensor hive through WLAN to navigate the robot to get the desired object.
% The cognitive robot navigation scenario is shown in Figure~\ref{fig_robot}, which is designed for the EEG data to drive PR2 robot to implement its serving task. 
Starting from the table near the kitchen, the PR2 robot receives action commands (as shown in Table~\ref{tab_command}) and walks forward until the specific position with the auxiliary of RFID tag. Then, the robot grasps the object, turns back and walks along the path to the table in living room and unlooses hands to put the beverage on the table.
The simulation result shows that the robot can 100\% precisely grasp and unloose object according to the path planned in the subject's mind. The simulation platform is in Gazebo toolbox and the robot controlling program is powered by Robot Operating System (ROS). This case randomly selects some EEG raw data from Subject 1 dataset as simulation inputs.
 % The PR2 mobile robot is employed for the advantage of smooth disk base and flexible hand design.

%michael: the link to IoT is very weak for both cases. It would be worthwhile to think how to make this link stronger. For example, in this second one, can is an IoT object with RFID tag to be identified by robot (I suppose that your robot does not use vision technique. If it does, hard to link to IoT object). 

\section{Discussion} % (fold)
\label{sec:discussion}
% This paper proposes a novel unified deep learning based framework for enabling human-thing cognitive interactivity. The efficiency and effectiveness are demonstrated by the experiments while the deployment feasible is demonstrated by the case studies. However, there are still 
Here we present several open challenges: 1) the experiment only contains 7 subjects limited by the practical conditions, a larger and more diverse dataset is necessary to illustrate the effects of the proposed model; 2) the SAM component with focal zone is designed to automatically explore the latent dimension sequence of the input EEG data, nevertheless, the employment of SAM increases the training time resulted from more iterations of the LSTM cell; 3) most importantly, the RS stage shuffles the order and replicate the number of input dimensions to discover the optimal order in order to the best performance, but the optimal order can not be guaranteed to appear after the RS, thus try more times if the classification result is unsatisfactory; 4) the WAS-LSTM exploits the spatial information among EEG channels, thus a number of channels are required to provide enough information. 
% Based on our experience, the minimum amount of EEG channel is 8 to keep a competitive performance. The aforementioned challenges will be addressed in the future work.

% ; \red{5) in the case study, we only implement the proposed model for classification, which is illustrated most power in the experiments, but haven't implemented other algorithms such as LSTM and GRU.

\section{Conclusion} % (fold)
\label{sec:conclusion}
We propose a unified deep learning framework to bridge 
%michael: give the full name of BCI. people do not have to go the first page to see what bci is if they forget.
Brain-Computer Interface 
and Internet of Things in order to enable cognitive interactivity. We propose 
%michael: again, better to avoid only short names in conclusion. Please give more information
WAS-LSTM 
to extract inter-dimension dependency among the input signal of the human brain activities which are selected by the selective attention mechanism. We conduct %local 
real-world experiments to evaluate the proposed framework and the results demonstrate that our model outperforms the state-of-the-art baselines. Furthermore, our experience in developing two 
%real-time 
case studies, namely the brain typing system and the cognitive robot, are reported in the paper. These case studies 
%to 
validate the feasibility of the proposed framework.

%

% %
% \begin{figure*}[!t]
% \centering
% \subfloat[Case I]{\includegraphics[width=2.5in]{ZhPicture.jpg}%
% \label{fig_first_case}}
% \hfil
% \subfloat[Case II]{\includegraphics[width=2.5in]{ZhPicture.jpg}%
% \label{fig_second_case}}
% \caption{Simulation results for the network.}
% \label{fig_sim}
% \end{figure*}

%

% Note that the IEEE does not put floats in the very first column
% - or typically anywhere on the first page for that matter. Also,
% in-text middle ("here") positioning is typically not used, but it
% is allowed and encouraged for Computer Society conferences (but
% not Computer Society journals). Most IEEE journals/conferences use
% top floats exclusively. 
% Note that, LaTeX2e, unlike IEEE journals/conferences, places
% footnotes above bottom floats. This can be corrected via the
% \fnbelowfloat command of the stfloats package.

% trigger a \newpage just before the given reference
% number - used to balance the columns on the last page
% adjust value as needed - may need to be readjusted if
% the document is modified later
%\IEEEtriggeratref{8}
% The "triggered" command can be changed if desired:
%\IEEEtriggercmd{\enlargethispage{-5in}}

% references section

% can use a bibliography generated by BibTeX as a .bbl file
% BibTeX documentation can be easily obtained at:
% http://mirror.ctan.org/biblio/bibtex/contrib/doc/
% The IEEEtran BibTeX style support page is at:
% http://www.michaelshell.org/tex/ieeetran/bibtex/
\bibliographystyle{IEEEtran}
\bibliography{IoTJ.bib}
\vspace{-10mm}
% \begin{IEEEbiography}{Xiang Zhang} % Lina Yao, Shuai Zhang, Salil, Micheal, BB, Yunhao Liu ?
\begin{IEEEbiography}
[{\includegraphics[width=1in,height=1.25in,clip,keepaspectratio]{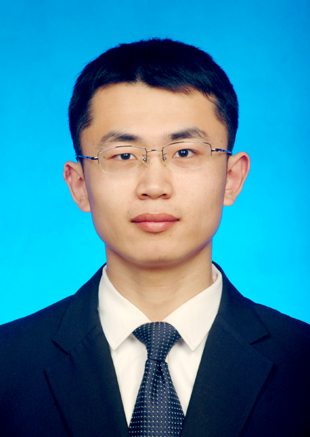}}]{Xiang Zhang} is currently a Ph.D. student (since 2016) at School of Computer Science and Engineering, University of New South Wales (UNSW). He received the Master degree (in 2016) from Harbin Institute of Technology (HIT), China. His research interests mainly in Deep learning, Brain-Computer Interface (BCI), Internet of Things (IoT), and Human Activity Recognition.
\end{IEEEbiography}
\vspace{-10mm}
\begin{IEEEbiography}
[{\includegraphics[width=1in,height=1.25in,clip,keepaspectratio]{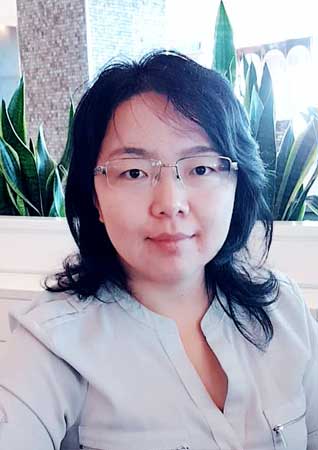}}]{Lina Yao} received the PhD degree in computer
science from the University of Adelaide, Australia.
She is currently a lecturer in the School of
Computer Science and Engineering, UNSW. Her
research interests lie in machine learning and data mining with applications to the Internet of Things, Brain-Computer Interface (BCI), information filtering and recommending, and
human activity recognition. She is a member of
the IEEE and the ACM.
\end{IEEEbiography}
\vspace{-10mm}
\begin{IEEEbiography}
[{\includegraphics[width=1in,height=1.25in,clip,keepaspectratio]{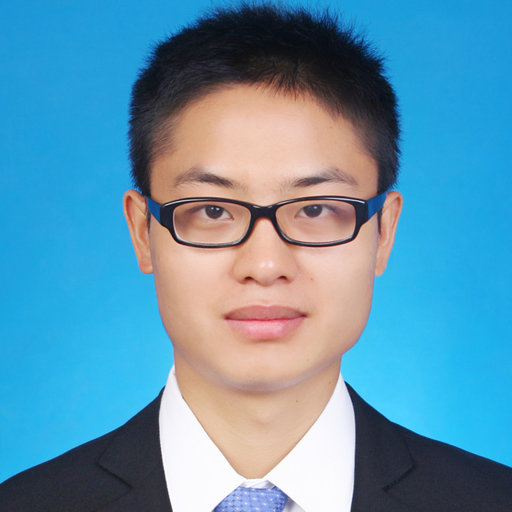}}]{Shuai Zhang}
 is a PhD student at the School of Computer Science and Engineering, University of New South Wales , as well as at Data61, CSIRO. He received a Bachelor degree from the School of Information Management, Nanjing University. His major research interests lie in the field of recommender systems, deep learning and internet of things. 
 He is a student member of the IEEE and ACM.
\end{IEEEbiography}
\vspace{-10mm}
\begin{IEEEbiography}
[{\includegraphics[width=1in,height=1.25in,clip,keepaspectratio]{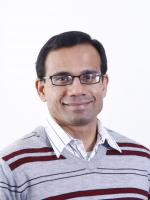}}]{Salil Kanhere}
received his Ph.D. from Drexel University. He is an associate professor in
the School of Computer Science and Engineering at UNSW.
His research interests include the Internet of Things, pervasive
computing, crowd sourcing, sensor networks, and security. He
has published 170 peer-reviewed articles and delivered over 20
tutorials and keynote talks. He is a Senior Member of ACM. He
is a recipient of the Humboldt Research Fellowship. 
\end{IEEEbiography}
\vspace{-10mm}
\begin{IEEEbiography}
[{\includegraphics[width=1in,height=1.25in,clip,keepaspectratio]{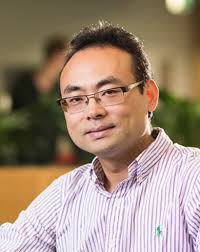}}]{Michael Sheng}
 is a full Professor and Head of Department of Computing at Macquarie University. Before moving to Macquarie, Michael spent 10 years at School of Computer Science, the University of Adelaide (UoA). Michael holds a PhD degree in computer science from the University of New South Wales (UNSW) and did his post-doc as a research scientist at CSIRO ICT Centre. From 1999 to 2001, Sheng also worked at UNSW as a visiting research fellow. 
 % Prior to that, he spent 6 years as a senior software engineer in industries.
 \end{IEEEbiography}
 \vspace{-10mm}
\begin{IEEEbiography}
[{\includegraphics[width=1in,height=1.25in,clip,keepaspectratio]{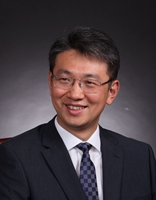}}]{Yunhao Liu}
received his BS degree in Automation Department from Tsinghua University, and an MA degree in Beijing Foreign Studies University, China. He received an MS and a Ph.D. degree in Computer Science and Engineering at Michigan State University, USA. Yunhao is now MSU Foundation Professor and Chairperson of Department of Computer Science and Engineering, Michigan State University, and holds Chang Jiang Chair Professorship (No Pay Leave) at Tsinghua University.\end{IEEEbiography}

\vfill

% if you will not have a photo at all:
% \begin{IEEEbiographynophoto}{John Doe}
% Biography text here.
% \end{IEEEbiographynophoto}

% % insert where needed to balance the two columns on the last page with
% % biographies
% %\newpage

% \begin{IEEEbiographynophoto}{Jane Doe}
% Biography text here.
% \end{IEEEbiographynophoto}

% You can push biographies down or up by placing
% a \vfill before or after them. The appropriate
% use of \vfill depends on what kind of text is
% on the last page and whether or not the columns
% are being equalized.

%\vfill

% Can be used to pull up biographies so that the bottom of the last one
% is flush with the other column.
%\enlargethispage{-5in}

% that's all folks
\end{document}